\documentclass[sigconf]{acmart}

\usepackage{color}
\usepackage{multirow}

\definecolor{formalshade}{rgb}{0.85,1,0.85}

\newcommand{\coloredbox}[3]{%
  \vspace{.25em}
  \noindent
  \setlength{\fboxsep}{5pt}%
  \colorbox{#1}{\parbox{\dimexpr\linewidth-2\fboxsep}{\textcolor{black}{\textbf{#2}} -- \textit{#3}}}%
  \vspace{.01em}
}
\AtBeginDocument{%
  \providecommand\BibTeX{{%
    \normalfont B\kern-0.5em{\scshape i\kern-0.25em b}\kern-0.8em\TeX}}}

\copyrightyear{2024}
\acmYear{2024}
\setcopyright{rightsretained}
\acmConference[WiseML '24]{Proceedings of the 2024 ACM Workshop on Wireless Security and Machine Learning }{May 31, 2024}{Seoul, Republic of Korea}
\acmBooktitle{Proceedings of the 2024 ACM Workshop on Wireless Security and Machine Learning (WiseML '24), May 31, 2024, Seoul, Republic of Korea}\acmDOI{10.1145/3649403.3656486}
\acmPrice{}
\acmISBN{979-8-4007-0602-8/24/05}

\begin{document}

\title[CANEDERLI]{CANEDERLI: On The Impact of Adversarial~Training and Transferability on CAN Intrusion Detection Systems}

\author{Francesco Marchiori}
\orcid{0000-0001-5282-0965}
\affiliation{%
 \institution{University of Padova}
 \streetaddress{Via Trieste, 63}
 \city{Padua}
 \country{Italy}}
\email{francesco.marchiori.4@phd.unipd.it}

\author{Mauro Conti}
\orcid{0000-0002-3612-1934}
\affiliation{%
 \institution{University of Padova}
 \streetaddress{Via Trieste, 63}
 \city{Padua}
 \country{Italy}}
\affiliation{%
 \institution{Delft University of Technology}
 \streetaddress{Mekelweg 4, 2628 CD}
 \city{Delft}
 \country{Netherlands}}
\email{mauro.conti@unipd.it}

\begin{abstract}
The growing integration of vehicles with external networks has led to a surge in attacks targeting their Controller Area Network (CAN) internal bus.
As a countermeasure, various Intrusion Detection Systems (IDSs) have been suggested in the literature to prevent and mitigate these threats.
With the increasing volume of data facilitated by the integration of Vehicle-to-Vehicle (V2V) and Vehicle-to-Infrastructure (V2I) communication networks, most of these systems rely on data-driven approaches such as Machine Learning (ML) and Deep Learning (DL) models.
However, these systems are susceptible to adversarial evasion attacks.
While many researchers have explored this vulnerability, their studies often involve unrealistic assumptions, lack consideration for a realistic threat model, and fail to provide effective solutions.

In this paper, we present \textbf{CANEDERLI} (\textbf{CAN} \textbf{E}vasion \textbf{D}etection \textbf{R}esi\textbf{LI}ence), a novel framework for securing CAN-based IDSs.
Our system considers a realistic threat model and addresses the impact of adversarial attacks on DL-based detection systems.
Our findings highlight strong transferability properties among diverse attack methodologies by considering multiple state-of-the-art attacks and model architectures.
We analyze the impact of adversarial training in addressing this threat and propose an adaptive online adversarial training technique outclassing traditional fine-tuning methodologies with F1 scores up to 0.941.
By making our framework publicly available, we aid practitioners and researchers in assessing the resilience of IDSs to a varied adversarial landscape.
\end{abstract}




\maketitle

\section{Introduction}
\label{sec:introduction}

The proliferation of advanced functionalities in modern vehicles necessitates an increased number of Electronic Control Units (ECUs).
As such, communication between these components becomes vital for ensuring the reliable operation of the vehicle's systems and features.
This heightened communication underscores the critical role of the Controller Area Network (CAN) bus in facilitating seamless interaction among ECUs.
Furthermore, the scope of communication extends beyond the confines of the vehicle itself, including interactions with external entities such as other vehicles (V2V) and infrastructures (V2I).
These communication protocols enable various functionalities, including cooperative driving, real-time traffic management, and advanced driver assistance systems~\cite{el2020vehicle}.

This heightened connectivity also increases potential security threats, prompting the need for robust Intrusion Detection Systems (IDSs).
These security tools are designed to monitor network or system activities for malicious activities or policy violations~\cite{rajbahadur2018survey}.
They analyze incoming network traffic, system logs, or other data sources and alert administrators or take action when they detect suspicious behavior or known attack patterns.
Due to the benefits offered by data-driven methodologies, Machine Learning (ML) and Deep Learning (DL) techniques have gained significant traction and are widely utilized in implementing IDSs~\cite{khraisat2019survey}.

The widespread usage of Artificial Intelligence (AI) for generating IDSs makes them vulnerable to adversarial attacks~\cite{goodfellow2014explaining}.
These attacks involve maliciously crafted input data that deceive ML and DL models into making incorrect predictions or classifications.
By exploiting vulnerabilities in the learning algorithms, adversaries can manipulate the behavior of IDSs and evade detection, potentially leading to system compromise or malfunction~\cite{qiu2020adversarial}.
As a result, robust defenses against adversarial attacks are essential to ensure the reliability and security of vehicle systems.
However, while extensive research has been conducted on adversarial attacks targeting IDSs of vehicle networks, these studies often require a set of assumptions that might not be realistic in real-world applications.
As such, while the effectiveness of these attacks is alarmingly high in most studies, their threat model might not reflect the capabilities of actual attackers.
Additionally, while some countermeasures have been proposed in the literature, their practicality and effectiveness against varied attacks are unclear.

\paragraph{Contribution.}
To address this gap in the literature on the practicality of adversarial attacks towards vehicle networks IDSs, we present \textbf{CANEDERLI}, the first framework for evaluating the impact of transferability and adversarial training in the context of CAN-based IDSs.
Our framework includes several model architectures and state-of-the-art attacks, allowing for a comprehensive evaluation of the adversarial impact.
In the pursuit of a realistic threat model, our attacks are generated in white-box and black-box scenarios, reflecting the capabilities of real-world attackers.
We incorporate adversarial training methodology based on fine-tuning procedures.
Additionally, we introduce \textit{adaptive online adversarial training}, which surpasses traditional techniques by preserving high accuracy and F1 score even during attacks, all while ensuring that the baseline performance of the models remains uncompromised.
By making our framework open-source, we allow for the complete customization of models and attacks, allowing researchers and practitioners to evaluate better the resilience of their IDSs.
Our contributions can be summarized as follows.
\begin{itemize}
    \item We propose a novel framework for evaluating the impact of adversarial attacks and securing IDSs.
    \item We propose an adversarial training technique outclassing fine-tuning-based methodologies.
    \item We evaluate our system on a real-world dataset.
    Our evaluation includes multiple state-of-the-art attacks and model architectures.
    \item We open-source our code at: \url{https://github.com/Mhackiori/CANEDERLI}.
\end{itemize}

\paragraph{Organization.}
The paper is organized as follows.
We analyze the literature on IDSs and their attacks in Section~\ref{sec:related}.
Our system and threat model are detailed in Section~\ref{sec:threat}.
Section~\ref{sec:methodology} delves into the methodology of our framework and the technical details of our contributions.
We evaluate our framework in Section~\ref{sec:evaluation} and provide valuable takeaways from our study in Section~\ref{sec:takeaways}.
Finally, Section~\ref{sec:conclusions} concludes our work.
\section{Related Works}
\label{sec:related}

Several IDSs have been recently proposed in the literature for vehicle networks.
This section focuses on data-driven approaches that leverage ML and DL models.
Starting from traditional on-road vehicles, the focus of IDSs has been the CAN bus.
This protocol was developed by Robert Bosch GmbH in the 1980s and has quickly become a standard in internal vehicle networks due to its convenience for safety-critical applications~\cite{canbus_standard}.
Most techniques for detecting intrusions or attacks in the CAN bus involve ML models, frequency-based methods, statistical-based methods, or hybrid approaches~\cite{lokman2019intrusion}.
For example, Kang et~al. were the first to utilize a semi-supervised Deep Neural Network (DNN) for this purpose~\cite{kang2016intrusion}.
While some works propose the use of simple linear networks~\cite{chalapathy2019deep}, others use more complex models such as Convolutional Neural Networks and Long Short Term Memory (LSTMs)~\cite{tariq2020cantransfer}.
Other approaches instead utilize the causality between data samples to predict the next value in a given sequence and evaluate divergence from the prediction~\cite{avalappampatty2015dynamic}.
Traffic frequency and statistics have also been used for this scope.
Indeed, authors could obtain high accuracy in detecting anomalies by analyzing the behavior of interacting ECUs and extracting statistical properties of traffic time series~\cite{song2016intrusion}.
The combination of these approaches yields the best results in the most varied scenarios, as ML models can leverage statistical and frequency data as features.
This also allows for the real-time deployment of these systems and the online processing of the traffic~\cite{weber2018embedded}.
With the increased connectivity of vehicles with other vehicles or infrastructures, the scope of IDSs has expanded to consider also V2V and V2I~\cite{aloqaily2019intrusion}.
As such, while IDSs can still be mounted on the internal networks, they need to consider additional threats from the external environment.
An even more challenging scenario is represented by Autonomous Vehicles (AVs), which, being equipped with several sensors and actuators, require heightened connectivity for their safe deployment.

While advantageous in terms of accuracy and complexity, the usage of ML and DL models for intrusion detection makes them vulnerable to adversarial attacks.
Adversarial attacks involve crafting malicious input data to deceive AI models into making incorrect predictions or classifications, potentially leading to system compromise or malfunction.
One class of adversarial techniques is evasion attacks, which entail crafting specific perturbations to input data to induce misclassification in the target model~\cite{goodfellow2014explaining}.
Another class is poisoning attacks, which manipulate the training data to compromise the model's performance at test time.
While the real-world implementation of these attacks requires a set of assumptions on the attacker's capability, their application towards vehicle-based IDSs has been studied in the literature~\cite{ayub2020model}.
However, given the restricted threat model of attacks toward vehicles, one property that needs investigation is adversarial transferability, i.e., the capability of adversarial examples crafted for one model to fool another~\cite{alecci2023dumb}.
While the properties of these attacks have been partly studied in the literature~\cite{zenden2023resilience}, their real-world impact and consequences remain uncertain.
This lack of clarity complicates the formulation of effective defenses against such attacks.
\section{System and Threat Model}
\label{sec:threat}

We now discuss the assumptions that characterize the system functionality and the attacker's capability.

\paragraph{System Model.}
IDSs require access to the vehicle network traffic to operate.
As such, the most straightforward implementation of the system is as an ECU connected directly to the CAN bus.
In this scenario, access to the encoding and decoding schema of the vehicle packets is not required, as intrusion detection can be performed at the bit level.
Furthermore, ECU IDSs can act as a filter and thus discard or flag malicious messages.
Alternatively, IDSs can be implemented in the cloud.
In this case, the vehicle should send the raw network traffic through an active external connection.
This last scenario also allows for implementing more complex model architectures, as ECUs might have limited resources available for computation.
Regardless of the implementation details, the IDS is constituted by a ML or DL model, taking in input single packet samples (or time windows, if using causal models) and producing an output flagging the packets as legitimate or malicious.
In the case of abnormal packets, the model can also discern what type of intrusion it detects.

\paragraph{Threat Model.}
In real-world scenarios, an attacker might aim to compromise the security and safety of a vehicle without being detected by the IDS.
For example, the CAN bus is notoriously vulnerable to Denial of Service (DoS) attacks, which can cause severe incidents if not promptly addressed~\cite{palanca2017stealth}.
As such, we assume the attacker can read the vehicle network traffic and inject messages into the bus.
This can be done through physical access to the On-Board Diagnostic (OBD) port or compromised connection toward other vehicles or infrastructures.
When injecting malicious data, the attacker must apply perturbations to the attack packets.
This perturbation needs to be substantial enough to avoid detection by the IDS but not too noticeable, as it might nullify the effect of the attack.
However, in real-world scenarios, attackers cannot access the target model.
This implies not being able to use its parameters and gradient for white-box attacks or using it as an oracle for crafting black-box attacks.
Furthermore, training and validation datasets are kept private from the IDS manufacturers to prevent poisoning attacks.
Thus, the attacker knows the classes (i.e., the attacks) labeled by the model but cannot access the source of the training data or its statistical distribution.
We formalize three scenarios to comprehensively study the attacker's capabilities under different assumptions.
\begin{itemize}
    \item \textit{White-Box (WB)}: the attacker can access the target vehicle data and target model.
    \item \textit{Gray-Box (GB)}: the attacker can access either the target vehicle data or the target model.
    \item \textit{Black-Box (BB)}: the attacker cannot access the target vehicle data or the target model.
\end{itemize}
In the gray-box and black-box scenarios, the attacker can still perform evasion attacks by using surrogate models.
These models have different architectures or are trained on other datasets.
However, these attacks will be effective only if they present high transferability properties.
\section{Methodology}
\label{sec:methodology}

\begin{figure*}[!htpb]
    \centering
    \includegraphics[width=\textwidth]{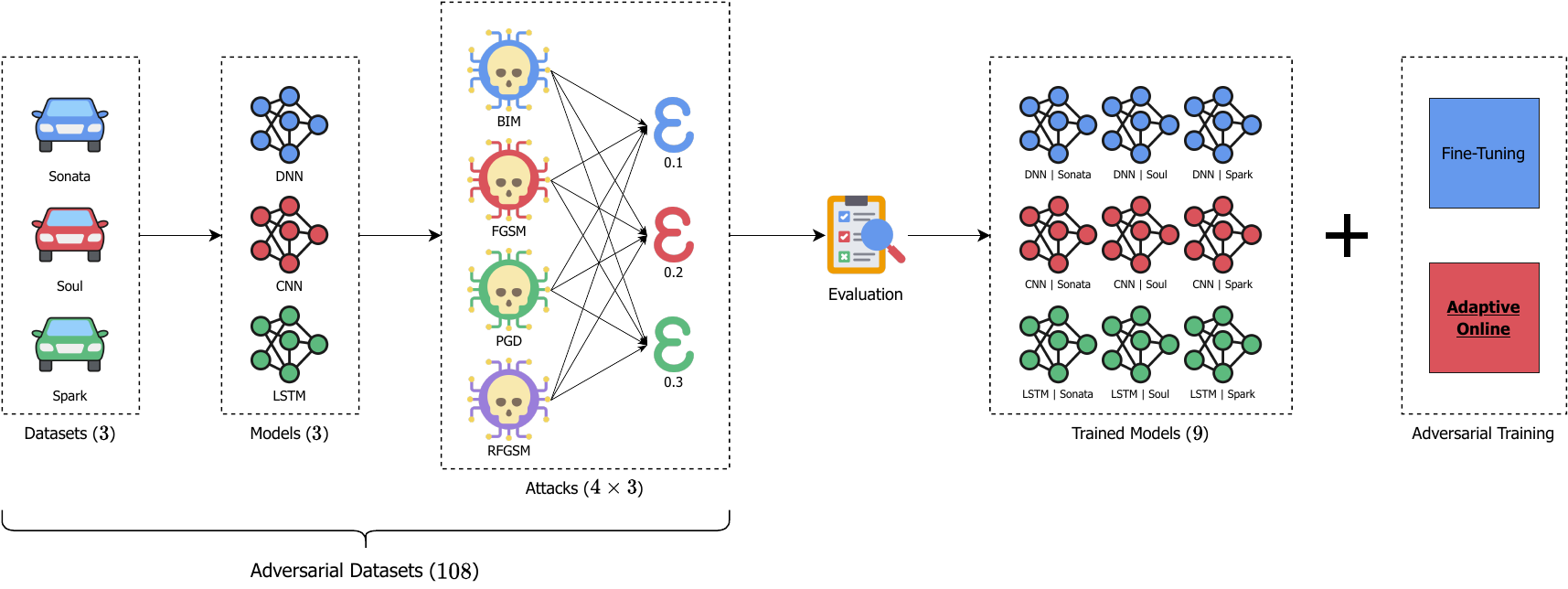}
    \caption{Framework overview.}
    \label{fig:framework}
    \Description[Framework overview.]{Framework indicating three vehicles used for the dataset, three models and four attacks generated at three different epsilon values, obtaining a total of 108 adversarial datasets. These datasets are then evaluated on nine different models (e.g., model 1 trained on dataset 1, model 2 trained on dataset 1), on which two different adversarial training techniques are applied.}
\end{figure*}

We now delve into the methodology of CANEDERLI, our adversarial transferability and training framework.
We discuss the models we use for testing and their parameters in Section~\ref{subsec:models}.
Section~\ref{subsec:attacks} overviews the evasion attacks we employ in our system.
In Section~\ref{subsec:training}, we show the adversarial training methods we use to defend our models and propose our novel technique.
An overview of our framework is shown in Figure~\ref{fig:framework}.

\subsection{Models}
\label{subsec:models}

We develop three models to evaluate the effectiveness and resilience of different DL architectures acting as an IDS: a linear DNN, a CNN, and an LSTM.

\paragraph{DNN}
We utilize the linear DNN due to its simplicity and effectiveness in capturing linear relationships within the data, making it suitable for straightforward classification tasks.
The model consists of two fully connected layers with ReLU activation functions in between.
The input size is specified by the number of features we use (which will be discussed in Section~\ref{subsec:dataset}), and the hidden layer size is set to 64.

\paragraph{CNN}
We choose the CNN architecture for its ability to extract spatial features from data, which may reveal patterns indicative of intrusion activities in the network traffic.
Its architecture incorporates a convolutional layer followed by a max-pooling layer.
The input is a 1D signal, and the convolutional layer is defined with a kernel size of 3 and 32 output channels.
After the convolution and pooling operations, the output is flattened and fed into a fully connected layer, which produces the final classification scores.

\paragraph{LSTM}
We select the LSTM architecture to leverage its capacity to capture temporal dependencies and causal relationships within sequential data, which aligns well with the time-series nature of traffic data in a vehicular network.
The model utilizes an LSTM layer for sequential data processing.
It accepts input sequences with a length specified by the number of features and generates hidden states with a size of 64.
The final classification is performed using a fully connected layer applied to the output of the last LSTM time step.

\subsection{Attacks}
\label{subsec:attacks}

In line with our threat model, we focus on evasion attacks since poisoning attacks are impossible without access to the training dataset.
By defining $x$ as the original sample, malicious samples can be created as $x^* = x + r$, where $r$ is the perturbation.
Most evasion attacks craft $r$ through an optimization process similar to the following.
\begin{equation}
\label{eq:perturbation}
    r = \arg \min_z f(x + z) \neq f(x).
\end{equation}
In this equation, $z$ is the variable under optimization, representing the perturbation added to the initial input $x$ to yield the perturbed input $x + z$.
We generate adversarial samples on the test set with Torchattacks~\cite{kim2020torchattacks}, a popular Python library for crafting different evasion attacks~\cite{wu2022dnd}.
We focus on the following types of attacks.

\begin{itemize}
    \item \textit{Basic Iterative Method (BIM)} --
    BIM is an iterative adversarial attack method that perturbs input data in small steps toward maximizing the model's loss.
    It aims to generate adversarial examples by iteratively applying small perturbations to input features~\cite{kurakin2018adversarial}.
    \item \textit{Fast Gradient Sign Method (FGSM)} --
    FGSM is a single-step adversarial attack method that computes the gradient of the loss function with respect to the input data and perturbs the input in the direction of the gradient sign.
    Despite its simplicity, FGSM often produces effective adversarial examples~\cite{goodfellow2014explaining}.
    \item \textit{Projected Gradient Descent (PGD)} --
    PGD is an iterative variant of the FGSM attack where the perturbations are constrained within a specified epsilon ball around the original data.
    By iteratively applying small perturbations while projecting them back onto the epsilon ball, PGD aims to generate strong adversarial examples~\cite{madry2017towards}.
    \item \textit{Randomized Fast Gradient Sign Method} --
    RFGSM is a variant of the FGSM attack that introduces randomness in the perturbation process.
    By adding random noise to the perturbation direction, RFGSM aims to enhance the transferability and robustness of adversarial examples~\cite{tramer2017ensemble}.
\end{itemize}

As anticipated in Section~\ref{sec:threat}, the attacker needs to carefully scale the perturbation to be applied to the sample to avoid drastically modifying the packet values.
For instance, in the case of the FGSM attack, malicious samples are generated as follows.
\begin{equation}
\label{eq:fgsm}
    x^* = x + \varepsilon \cdot sign(\nabla_xJ(\theta, x, y)).
\end{equation}
Here, $\epsilon$ is the scaling factor, $J$ is the loss function, $\theta$ represents the model parameters, and $y$ is the ground truth for the input $x$.
Higher $\epsilon$ can yield a higher Attack Success Rate (ASR), while lower $\epsilon$ reduce the perceptibility of the attack.
To provide a comprehensive analysis of this tradeoff, we generate all attacks at three different maximum $\epsilon$ values: $0.1$, $0.2$, and $0.3$

\subsection{Adversarial Training}
\label{subsec:training}

One of the most effective approaches for defending against adversarial attacks is adversarial training~\cite{bai2021recent}.
This procedure involves including adversarial attacks at training time to enhance the model's robustness.
We use two different techniques for adversarial training.
The first involves fine-tuning the pre-trained model on adversarial samples.
While this approach has the advantage of being suitable to any trained model, it has the drawback of losing baseline performance.
We propose \textit{adaptive online adversarial training} to fix this issue.
For each training epoch and batch, we first train the model on the legitimate inputs, then generate each attack in the same batch, evaluate them on the model, and backpropagate the loss (\textit{``online''} as attacks are dynamically generated at each epoch)~\cite{efatinasab2024faultguard}.
Furthermore, to reflect the increasing capabilities of the model as training commences, each attack is scaled with an increasing $\epsilon$ (\textit{``adaptive''} as $\epsilon$ adapts to the model training).
In particular, at each epoch $i$, attacks are scaled as follows.
\begin{equation}
\label{eq:eps}
    \epsilon_i = \epsilon_{max} \frac{i}{num\_batches}. 
\end{equation}
This dynamic training approach confronts the model with progressively challenging adversarial examples, compelling continuous enhancement of its robustness.
\section{Evaluation}
\label{sec:evaluation}

We now provide an experimental evaluation of our framework.
First, in Section~\ref{subsec:dataset}, we give details on the used dataset and the extracted features.
In Section~\ref{subsec:baselineevaluation}, we evaluate our models in an adversary-free scenario, while in Section~\ref{subsec:attacksevaluation}, we evaluate the efficacy of our attacks.
Section~\ref{subsec:advtrevaluation} evaluates and compares the proposed adversarial training techniques.

\subsection{Dataset}
\label{subsec:dataset}

For the evaluation, we use the Survival dataset~\cite{han2018anomaly}.
This dataset focuses on three attack scenarios drastically affecting vehicle functions.
\begin{itemize}
    \item \textit{Flooding} --
    Being a multi-master network, the CAN bus manages collisions through arbitration.
    Thus, messages with higher priority (i.e., lower ID values) can overwrite packets with lower priority.
    This opens the possibility of DoS attacks, where attackers inject messages with low ID values to void the vehicle's functionalities.
    \item \textit{Fuzzy} --
    Fuzzing is a testing technique used in software development to find vulnerabilities or bugs by injecting random or anomalous inputs.
    This attack uses the same principle for malicious purposes by injecting packets with random IDs.
    \item \textit{Malfunction} --
    This attack focuses on specific IDs and overwriting their payload with different values from the original.
    This has the effect of abnormal vehicle behavior.
\end{itemize}

The dataset was collected by performing these attacks in three vehicles: HYUNDAI YF Sonata, KIA Soul, and CHEVROLET Spark.
This produces three different datasets, one for each vehicle.
The collected traffic comprises CAN bus packets containing their timestamp, ID, payload, and Data Length Code (DLC).
We pre-process the dataset by converting the payload from hexadecimal to binary, thus increasing the number of features.
Furthermore, we use the timestamp to compute intervals between messages with the same ID.
This way, detecting injected messages becomes more straightforward with constant upload speeds.
We label each packet with four possible values (three attacks and legitimate traffic).
Finally, we balance our dataset with undersampling to avoid bias due to the data distribution.
This ensures that, for each dataset, the number of packets for each label is the same.

\subsection{Baseline Evaluation}
\label{subsec:baselineevaluation}

We now evaluate the baseline performance of our IDSs.
Since we tune our model hyperparameters offline, we divide our dataset in training (80\% of the dataset) and testing (20\% of the dataset).
We train each model on each dataset, obtaining $3 \times 3 = 9$ trained models.
Each model is trained for 30 epochs, using Adam as the optimizer with a learning rate of $0.001$ and using cross entropy as the loss function.
Results are shown in Table~\ref{tab:baseline}.
All models obtain results close to perfection on all tasks.
While some datasets appear to be easier than others (e.g., Sonata), accuracy and F1 score are high enough on all vehicles to provide low false positive and false negative rates.

\begin{table}[!htpb]
  \centering
  \small
  \caption{Baseline performance of the models.}
  \label{tab:baseline}
  \begin{tabular}{l|cc|cc|cc}
    \hline
    \multirow{2}{*}{\textbf{Model}} & \multicolumn{2}{c|}{\textbf{Sonata}} & \multicolumn{2}{c|}{\textbf{Soul}} & \multicolumn{2}{c}{\textbf{Spark}} \\ \cline{2-7}
    & \textbf{Acc} & \textbf{F1} & \textbf{Acc} & \textbf{F1} & \textbf{Acc} & \textbf{F1} \\
    \hline
    \textbf{DNN} & 1.000 & 1.000 & 0.995 & 0.995 & 0.997 & 0.997 \\
    \textbf{CNN} & 0.999 & 0.999 & 0.993 & 0.993 & 0.997 & 0.997 \\
    \textbf{LSTM} & 1.000 & 1.000 & 0.995 & 0.995 & 0.995 & 0.995 \\
    \hline
  \end{tabular}
\end{table}

\subsection{Attacks Evaluation}
\label{subsec:attacksevaluation}

To assess the effectiveness of our attacks, we evaluate the accuracy and F1 scores of the models when tested on adversarial datasets.
Inspired by~\cite{alecci2023dumb}, we generate samples for each attack at each $\epsilon$ value for each model, obtaining $4 \times 3 \times 9 = 108$ adversarial datasets.
Each of them is then evaluated on each model, obtaining $108 \times 9 = 972$ evaluations.
We show the results in Table~\ref{tab:attacks-m}, where we split our evaluation based on the scenarios detailed in Section~\ref{sec:threat}.
The mean F1 score drop in all scenarios is significant as these attacks can completely disrupt the system's functionality.
Furthermore, we notice strong transferability properties as attacks in the gray-box and black-box scenarios are also effective.
This is due to the presence of attacks generated on different vehicle types, of which the characteristics are unknown to the legitimate system.
We notice that the CNN model appears to be the most resilient.
Instead, the types of attacks don't influence the score as much (Table~\ref{tab:attacks-a}).

\begin{table}[!htpb]
  \centering
  \small
  \caption{Average model' performance on adversarial datasets.}
  \label{tab:attacks-m}
  \begin{tabular}{l|c|c|c}
        \hline
        \multirow{2}{*}{\textbf{Model}} & \multicolumn{3}{c}{\textbf{F1 Score}} \\ \cline{2-4}
        & \textbf{WB} & \textbf{GB} & \textbf{BB} \\
        \hline
        \textbf{DNN} & 0.257 & 0.253 & 0.246  \\
        \textbf{CNN} & 0.462 & 0.440 & 0.374 \\
        \textbf{LSTM} & 0.254 & 0.294 & 0.283 \\
        \hline
      \end{tabular}
\end{table}
\begin{table}[!htpb]
  \centering
  \small
  \caption{Average attacks' performance.}
  \label{tab:attacks-a}
      \begin{tabular}{l|c|c|c}
        \hline
        \multirow{2}{*}{\textbf{Attack}} & \multicolumn{3}{c}{\textbf{F1 Score}} \\ \cline{2-4}
        & \textbf{WB} & \textbf{GB} & \textbf{BB} \\
        \hline
        \textbf{BIM} & 0.362 & 0.380 & 0.356 \\
        \textbf{FGSM} & 0.295 & 0.284 & 0.251 \\
        \textbf{PGD} & 0.285 & 0.293 & 0.265 \\
        \textbf{RFGSM} & 0.356 & 0.361 & 0.333 \\
        \hline
      \end{tabular}
\end{table}

\subsection{Adversarial Training Evaluation}
\label{subsec:advtrevaluation}

To defend against these threats, the best course of action is performing adversarial training on the target model.
First, we evaluate a traditional adversarial learning paradigm, consisting of fine-tuning the pre-trained model on an adversarial dataset.
Secondly, we evaluate our adaptive online adversarial learning technique.
The results are shown in Table~\ref{tab:training-ft} and Table~\ref{tab:training-ao}.
Fine-tuning-based adversarial learning methodologies can effectively defend against white-box attacks.
However, one major drawback makes implementing this technique impossible in real-world scenarios: significant drops in baseline performance.
As such, adversarially trained models lose their original accuracy and cannot perform in legitimate scenarios.
Instead, our proposed online adversarial learning method shows similar F1 scores when under attack but maintains the baseline performance of the original models.
Thus, our method balances performance and resilience against adversarial attacks.

\begin{table}[!htpb]
  \centering
  \small
  \caption{Fine-tuning-based adversarial training performance.}
  \label{tab:training-ft}
  \begin{tabular}{l|c|c|c|c}
        \hline
        \multirow{2}{*}{\textbf{Model}} & \multirow{2}{*}{\textbf{Clean}} & \multicolumn{3}{c}{\textbf{F1 Score}} \\ \cline{3-5}
        & & \textbf{WB} & \textbf{GB} & \textbf{BB} \\
        \hline
        \textbf{DNN} & 0.376 & 0.956 & 0.808 & 0.679  \\
        \textbf{CNN} & 0.559 & 0.997 & 0.782 & 0.692 \\
        \textbf{LSTM} & 0.367 & 0.963 & 0.804 & 0.661 \\
        \hline
      \end{tabular}
\end{table}

\begin{table}[!htpb]
  \centering
  \small
  \caption{Adaptive online adversarial training performance.}
  \label{tab:training-ao}
      \begin{tabular}{l|c|c|c|c}
        \hline
        \multirow{2}{*}{\textbf{Model}} & \multirow{2}{*}{\textbf{Clean}} & \multicolumn{3}{c}{\textbf{F1 Score}} \\ \cline{3-5}
        & & \textbf{WB} & \textbf{GB} & \textbf{BB} \\

        \hline
        \textbf{DNN} & 0.991 & 0.936 & 0.796 & 0.671 \\
        \textbf{CNN} & 0.996 & 0.880 & 0.741 & 0.621 \\
        \textbf{LSTM} & 0.998 & 0.941 & 0.808 & 0.673 \\
        \hline
      \end{tabular}
\end{table}
\section{Takeaways}
\label{sec:takeaways}

Our evaluation underscores the threat that evasion attacks pose to CAN-based IDSs.
Thus, we now discuss and summarize the main takeaway messages we identify for proposing secure implementations in real-world scenarios.

\coloredbox{formalshade}{Takeaway 1}{A detailed definition of the system and threat models is necessary for thoroughly evaluating the scope of the threat.}

Different implementations of an IDS involve different assumptions on its functioning and the adversary knowledge.
As highlighted by our results, the effect of attacks and defense measures highly depends on the application scenarios.
Thus, knowing the attackers' capabilities when designing the system can significantly improve the effectiveness of the implemented countermeasures.

\coloredbox{formalshade}{Takeaway 2}{Black-box attacks can be as threatening as white-box attacks, as using surrogate models is an effective solution for the attacker.}

While models behave differently based on their architectures and the datasets they are trained on, attack behavior is consistent across different scenarios.
For example, when performing a DoS attack, the best strategy is to flood the traffic with packets with low IDs.
As such, while regular traffic has different properties based on the encoding and the ECUs that constitute the vehicle, IDSs are trained to identify specific patterns used when also other vehicles are under attack.
This makes this class of evasion attacks highly transferable.

\coloredbox{formalshade}{Takeaway 3}{IDSs security should be tackled during the design process, as adversarial fine-tuning strategies might not be efficient.}

Even though IDSs are designed for security purposes, their security is paramount for ensuring their effectiveness.
As such, efficient adversarial training techniques are essential to defend against white-box and black-box evasion attacks.
However, as our analysis shows, fine-tuning techniques drastically lower their accuracy in baseline performance.
Therefore, our proposed adaptive online adversarial training is preferable.
A detailed definition of the threat model is necessary to effectively train the models, as knowing what attacks the models are most probably encountering can increase their accuracy both in baseline performance and under attack.
\section{Conclusions}
\label{sec:conclusions}

Modern vehicles rely on numerous ECUs and robust communication through the CAN bus.
Furthermore, communication extends to external entities such as other vehicles and infrastructures.
However, it also introduces security risks, necessitating the implementation of IDSs.
These systems usually leverage data-driven approaches, making them vulnerable to adversarial attacks.
Unfortunately, the current literature on these attacks often lacks realistic assumptions and effective countermeasures, underscoring the need for further investigation and solutions.

\paragraph{Contribution.}
CANEDERLI addresses the gap in research on adversarial attacks on vehicle network IDSs.
We introduce a framework for evaluating transferability and adversarial training impact, incorporating diverse model architectures and attacks.
Our framework ensures realism by considering white-box and black-box scenarios, offering adaptive online adversarial training, and preserving model performance under attack.
We open-source our framework for customization and IDS resilience assessment.

\paragraph{Future Works.}
Future research directions include conducting granular evaluations by examining various epsilon values in evasion attacks.
Furthermore, more intricate model architectures and diverse datasets are worth exploring.
This can also include the addition of diverse intrusion methodologies and attack techniques.
These efforts aim to improve the resilience of IDSs against sophisticated adversarial threats, ensuring robust protection in dynamic environments.

\bibliographystyle{ACM-Reference-Format}
\balance
\bibliography{references}

\end{document}